\documentclass[prl,nofootinbib,floats,aps,twocolumn,tightenlines,superscriptaddress,amsmath,amssymb]{revtex4-1}

\usepackage[english]{babel}
\usepackage{graphicx}
\usepackage{dcolumn}
\usepackage{bm}
\usepackage{verbatim}
\usepackage{mathrsfs}
\usepackage{cancel}
\usepackage{epstopdf}
\usepackage{color}
\usepackage{tikz}
\usetikzlibrary{arrows}
\usetikzlibrary{decorations.pathreplacing}

\DeclareMathOperator{\Tr}{Tr}

\DeclareMathOperator{\sgn}{sgn}

\DeclareMathOperator{\Realpart}{Re}

\newcommand{\ircutoff}{{\Lambda}}

\newcommand{\sfx}{{\sf x}}
\newcommand{\ket}[1]{\left| {#1} \right\rangle}
\newcommand{\bra}[1]{\left\langle {#1} \right|}

\newcommand{\ii}{\mathrm{i}}

\newcommand{\proj}[2]{\left| {#1} \right\rangle\!\left\langle {#2} \right|}

\usepackage[usenames,dvipsnames]{pstricks}
\usepackage{epsfig}
\usepackage{pst-grad} 
\usepackage{pst-plot} 

\usepackage{hyperref}

\usepackage{verbatim}

\begin{document}


\title{(1+1)D calculation provides evidence that quantum entanglement
survives a firewall}

\author{Eduardo Mart\'{i}n-Mart\'{i}nez}
\affiliation{Institute for Quantum Computing, University of Waterloo, Waterloo, Ontario, N2L 3G1, Canada}
\affiliation{Department of Applied Mathematics, University of Waterloo, Waterloo, Ontario, N2L 3G1, Canada}
\affiliation{Perimeter Institute for Theoretical Physics, Waterloo, Ontario, N2L 2Y5, Canada}

\author{Jorma Louko}
\affiliation{School of Mathematical Sciences, University of Nottingham, Nottingham NG7 2RD, UK}

\begin{abstract}
We analyze how pre-existing entanglement between two Unruh-DeWitt
particle detectors evolves when one of the detectors falls through a
Rindler firewall in (1+1)-dimensional Minkowski space. The firewall
effect is minor and does not wash out the detector-detector
entanglement, in some regimes even preserving the entanglement better
than Minkowski vacuum. The absence of cataclysmic events should continue
to hold for young black hole firewalls. A firewall's prospective ability
to resolve the information paradox must hence hinge on its detailed
gravitational structure, presently poorly understood.
\end{abstract}

\date{Revised June 2015}

\maketitle

\textit{\textbf{Introduction.---}}
If black hole evaporation preserves unitarity, 
it has been argued from preservation of correlations that 
the horizon of a shrinking black hole must develop
a singularity even when the evaporation 
is still slow and the black hole remains macroscopic 
\cite{Mathur:2009hf,braunstein-et-al,Almheiri:2012rt}
(for a selection of debate and reviews see 
\cite{Susskind:2013tg,Almheiri:2013hfa,Page:2013mqa,marolf-polchinski-duality-bhint,Hotta:2013clt,Almheiri:2013wka,Harlow:2014yka,Hotta:2015yla}).
Modeling the gravitational aspects of this 
proposed singularity has remained elusive. 
For instance, the emergence of a
Planck scale shell near the horizon of an 
astrophysical black hole 
seems to be in tension with the 
gravitational dynamics predicted by general relativity~\cite{Abramowicz}. 
Nevertheless, the nongravitational aspects of the ``firewall'' 
version of the singularity \cite{Almheiri:2012rt} 
can be modeled
with a quantum field in Minkowski spacetime: 
a state in which correlations across a Rindler horizon 
are severed can be written down by hand~\cite{Louko2014},  
mimicking the severing that the firewall argument of 
\cite{Almheiri:2012rt} posits to develop 
dynamically during black hole evaporation. 
This severing of correlations
has strong similarities to that which   
ensues on the sudden insertion 
of a reflective wall in a spacetime~\cite{Brown:2014qna,Brown2015}.
The Rindler firewall state 
can be studied by usual quantum field theory techniques, 
and the conclusions should apply to young gravitational firewalls 
where the backreaction on the metric is still small. 

In this letter we analyse the correlations between two particle detectors 
when one of them falls through a Rindler firewall. 
From the quantum field theory side, this question is motivated by the fact that 
any measurements of quantum fields are done 
through material particle detectors, and we may employ the Unruh-DeWitt detector model 
\cite{Unruh,DeWitt} to capture the essential aspects of 
interactions between atoms and the electromagnetic field~\cite{Wavepackets,Alvaro}. 
From the firewall side, focusing on the correlations between two detectors is motivated by 
the central role of quantum correlations in the firewall argument~\cite{Almheiri:2012rt}. 
The response of a single detector falling through the Rindler 
firewall is known to be sudden but finite~\cite{Louko2014}. For 
two detectors that are initially correlated, 
interaction with the quantum field will decohere the two detectors: 
might this decoherence be drastically enhanced when one of the 
detectors goes through a Rindler firewall? 
A positive answer could be seen as indirect support of the 
firewall argument as given in~\cite{Almheiri:2012rt}. 

Our main conclusion runs contrary to these expectations. The Rindler 
firewall turns out to have only a modest effect on the 
decoherence between the two detectors, and 
in certain regions of the parameter space the firewall even 
preserves the entanglement between the two detectors better 
than Minkowski vacuum. A~firewall's prospective capability
to resolve the black hole information paradox must hence 
hinge on its detailed late time gravitational structure, 
at present poorly understood. 

\textit{\textbf{Formalism: evolving an inertial detector pair.---}} 
We consider a pair of Unruh-DeWitt detectors \cite{Unruh,DeWitt}
coupled to a real scalar field $\phi$ and  
moving inertially in Minkowki spacetime without 
relative velocity~\cite{Reznik2005}. 
(For an inertial detector and an accelerated detector, 
see, e.g.,~\cite{Lin:2008jj,Ostapchuk:2011ud,Lin:2015ama}.) 
We evolve the system with respect to the Minkowski time~$t$ 
in a Lorentz frame in which the two detectors are at rest, 
and we denote $t$ by $\tau$ as it coincides with the 
detectors' proper time. 
The interaction picture Hamiltonian is 
\begin{align}
H=\sum_\nu \lambda_\nu \chi_\nu(\tau)\mu_\nu(\tau)\phi\bigl(\sfx_\nu(\tau)\bigr)
\ , 
\label{eq:H-int-def}
\end{align} 
where the index $\nu$ labels the two detectors, 
$\sfx_\nu(\tau)$ are their worldlines, 
$\mu_\nu(\tau)$ are the monopole moment operators 
and $\lambda_\nu$ are the coupling constants. 
The real-valued switching functions 
$\chi_\nu$ specify how the interaction is turned on and off. 
We assume that $\chi_\nu$ either have 
compact support or have sufficiently strong falloff properties 
for the system to be treatable as asymptotically uncoupled 
in the distant past and future. 

If the initial state (= density matrix) 
of the system is~$\rho_0$, 
the final state after the interaction has ceased is 
$\rho_T = U \rho_0 U^\dagger$, 
where $U$ is the interaction picture time evolution operator. 
Assuming that each $\lambda_\nu$ is proportional to a 
formal perturbative parameter $\lambda$, 
$U$ has the Dyson expansion 
$U = U^{(0)} + U^{(1)} + U^{(2)} 
+ \mathcal{O}(\lambda^3)$ 
where $U^{(0)} = \openone$ 
and
\begin{align} 
U^{(1)}\!\!
=\! 
- \ii\!\!\int_{-\infty}^{\infty}\!\!\!\!\!\! \text{d} \tau 
\, H(\tau)
, \;
U^{(2)}\! =\! 
- \!\!\int_{-\infty}^{\infty}
\!\!\!\!\!\text{d}\tau \!\!\int_{-\infty}^{\tau}\!\!\!\!\!\text{d}\tau'
H(\tau) H(\tau')
\ . 
\label{eq:U2-def}
\end{align}
Hence 
$\rho_T  = \rho_0+\rho_T^{(1)}+\rho_T^{(2)}+\mathcal{O}(\lambda^3)$, where 
\begin{subequations}
\begin{align}
\label{eq:rho1}\rho_T^{(1)}
&=
U^{(1)}\rho_0+\rho_0{U^{(1)}}^\dagger 
\ , 
\\
\label{eq:rho2}\rho_T^{(2)}
&=
U^{(1)}\rho_0{U^{(1)}}^\dagger+U^{(2)}\rho_0+\rho_0{U^{(2)}}^\dagger
\ . 
\end{align}
\end{subequations}
When the initial state has the form 
$\rho_0=\rho_{\text{d},0}\otimes \rho_{\phi,0}$, 
where $\rho_{\text{d},0}$ and $\rho_{\phi,0}$ 
are respectively the initial state of the 
two-detector subsystem and the initial state of the field, 
and assuming that $\rho_{\phi,0}$ satisfies 
\begin{align}
\label{order1cond}
\Tr_{\phi}\bigl(\phi({\sfx})\rho_{\phi,0}\bigr)=0
\ , 
\end{align} 
we find that the final state of the two-detector subsystem is 
\begin{subequations}
\label{eq:rho-d-T-expanded}
\begin{align}
\rho_{\text{d},T} 
&= \Tr_\phi(\rho_T)
= 
\rho_{\text{d},0} + \rho_{\text{d},T}^{(2)}
+ 
\mathcal{O}(\lambda^3) 
\ , 
\label{orders}
\\
\rho_{\text{d},T}^{(2)}
& =\sum_{\nu,\eta}\lambda_\nu\lambda_\eta\bigg[ \int_{-\infty}^{\infty}\!\!\!\text{d}\tau \int_{-\infty}^{\infty}\!\!\!\text{d}\tau' \, 
\chi_\nu(\tau')\chi_\eta(\tau) 
\notag
\\
&\hspace{13ex}
\times\mu_\nu(\tau') \rho_{\text{d},0} \mu_\eta(\tau) \, W[\sfx_\eta(\tau),\sfx_\nu(\tau')]
\notag
\\
&\hspace{4ex}
- \int_{-\infty}^{\infty}\!\!\!\text{d}\tau \int_{-\infty}^{\tau}\!\!\!\text{d}\tau' \, 
\chi_\nu(\tau)\chi_\eta(\tau') 
\notag
\\
&\hspace{13ex}
\times\mu_\nu(\tau) \mu_\eta(\tau') \rho_{\text{d},0} \, 
W[\sfx_\nu(\tau),\sfx_\eta(\tau')]
\notag
\\
&\hspace{4ex}
- \int_{-\infty}^{\infty}\!\!\!\text{d}\tau \int_{-\infty}^{\tau}\!\!\!\text{d}\tau' \, 
\chi_\nu(\tau)\chi_\eta(\tau') 
\notag
\\
&\hspace{13ex}
\times\rho_{\text{d},0} \mu_\eta(\tau') \mu_\nu(\tau) \, 
W[\sfx_\eta(\tau'),\sfx_\nu(\tau)]\bigg] 
\label{eq:rho-d-T-2}
\end{align}
\end{subequations}
where $W[\sfx_\nu(\tau),\sfx_\eta(\tau')]$ denotes 
the pullback of the Wightman function on the detectors' worldlines, 
\begin{align}
W[\sfx_\nu(\tau),\sfx_\eta(\tau')] = \Tr_{\phi}
\bigl(\phi\bigl({\sfx_\nu(\tau)}\bigr)\phi\bigl({\sfx_\eta(\tau')}\bigr)\rho_{\phi,0}\bigr)
\ . 
\end{align}

\textit{\textbf{Detectors with a Rindler firewall.---}}
We now specialise to $(1+1)$-dimensional Minkowski spacetime, 
$ds^2 = - dt^2 + dx^2 = - du\,dv$, 
where $u = t-x$ and $v=t+x$. 

We take $\phi$ to be massless and $\rho_{\phi,0}$ to be the Rindler
firewall state described in~\cite{Louko2014}. The one-point function of 
$\rho_{\phi,0}$ satisfies~\eqref{order1cond}, as follows by extending the
Wightman function discussion given in
\cite{Louko2014} to the one-point function. 
The Wightman function of $\rho_{\phi,0}$ is 
\begin{align}
W_F(\sfx,\sfx') &= 
\Tr_{\phi}\bigl(\phi({\sfx}) \phi({\sfx}')\rho_{\phi,0}\bigr)
\notag 
\\
&= 
W_0(\sfx,\sfx')+ \Delta W(\sfx,\sfx')
\ , 
\end{align}
where $W_0$ is the Wightman function in the Minkowki vacuum $\proj{0}{0}$ 
and $\Delta W$ is the correction due to the firewall. 
For $W_0$ we have 
\begin{align}
W_0(\sfx,\sfx') 
=\frac{-1}{4\pi}
\log \! 
\left[
\ircutoff^2 ( \epsilon + \ii \Delta u) ( \epsilon + \ii \Delta v) 
\right] , 
\label{eq:mzero-Wightman-Mink}
\end{align}
where $\Delta u = u - u'$, $\Delta v = v - v'$, 
the positive constant $\ircutoff$ 
is an infrared cutoff, the logarithm takes its principal branch and 
$\epsilon \to 0_+$. 
The full expression for $\Delta W(\sfx,\sfx')$ is lengthy but reduces for 
$v>0$ and $v'>0$ to 
\begin{align}
\Delta W(\sfx,\sfx')
&= \frac{1}{4\pi}\Big[\Theta (u) \theta (-u')+\theta (-u) \theta (u')\Big]
\notag
\\
& \hspace{3ex}
\times \left(\log (\Lambda  \left| u-u'\right| )+\ii\frac{\pi}{2}  \sgn(u-u')\right).
\label{eq:mzero-Wightman-corr}
\end{align}

In words, \eqref{eq:mzero-Wightman-Mink} and \eqref{eq:mzero-Wightman-corr}
show that when $\sfx$ and $\sfx'$ are to the future of the left-going Rindler horizon $t=-x$ 
but on opposite sides of the right-going Rindler horizon $t=x$, 
$W_F(\sfx,\sfx')$ is missing the contribution from the right-moving part of the field. 
This absence of correlations across the Rindler horizon models the 
absence of correlations that is argued to develop 
dynamically in an evaporating black hole spacetime~\cite{Almheiri:2012rt}. 

\begin{figure}
\includegraphics[width=0.38\textwidth]{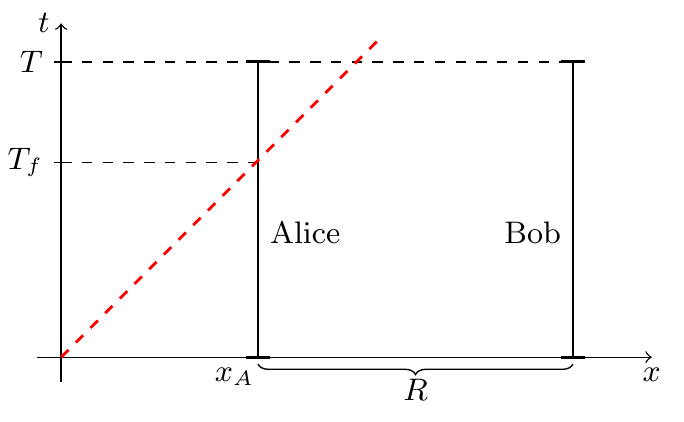}
\caption{Spacetime diagram of the two-detector system with the Rindler firewall. 
The dashed line at $t=x$ is the firewall. The 
solid lines are the worldlines of the Alice detector and the Bob detector, 
switched on at $t=0$ and off at $t=T>0$. 
Alice crosses the firewall during the detectors' operation 
(at $t=T_f=x_A$ in the diagram) but Bob does not.} 
\label{fig:stdiagram}
\end{figure}

For the detectors in the presence of the firewall, we 
take the worldline of detector $A$ (Alice) to be at $x=x_A>0$ and
the worldline of detector $B$ (Bob) to be at $x=x_A+R$, where $R>0$ is
the spatial separation.  The detectors are switched on at $t=0$, and
they are switched off at a time when Alice has already crossed the
firewall at $t = x$ but Bob has not, as shown in
Figure~\ref{fig:stdiagram}.

We ask: If Alice and Bob are initially entangled, 
how does Alice's crossing the firewall affect this entanglement? 

\textit{\textbf{Methods.---}}
We assume each detector to be a two-level system. We denote the 
respective energy gaps by~$\Omega_\nu$, 
the ground states by $\ket{g_\nu}$ and the excited states by~$\ket{e_\nu}$. 
The monopole moment operators are then 
$\mu_\nu(\tau) = \sigma_\nu^+ e^{\ii\Omega_\nu\tau} + \sigma_\nu^- e^{-\ii\Omega_\nu\tau}$, 
where the nonvanishing matrix elements of the raising and lowering operators 
$\sigma_\nu^\pm$ are 
$\bra{e_\nu} \sigma_\nu^+ \ket{g_\nu} = \bra{g_\nu} \sigma_\nu^- \ket{e_\nu} = 1$. 

For each of the the individual detectors we may introduce 
a two-by-two matrix representation in which 
(suppressing the detector index) 
\begin{align}
\ket{g}=\left (\!
\begin{array}{c}
1  \\
 0 \\
\end{array}
\!\right ),
\ 
\ket{e}=\left (\!
\begin{array}{c}
0  \\
1 \\
\end{array}
\!\right ),
\ 
\mu(\tau) = 
\left (\!
\begin{array}{c c}
0 & e^{-\ii \Omega \tau} \\
e^{\ii \Omega \tau} & 0 \\
\end{array}
\!\right )
\, . 
\label{eq:mu-matrixrep}
\end{align} 
For the two-detector system we employ the 
Kronecker product representation in which 
\begin{equation}
\ket{gg}=\begin{pmatrix}
1\\
0\\
0\\
0
\end{pmatrix},
\ket{eg}=\begin{pmatrix}
0\\
1\\
0\\
0
\end{pmatrix},
\ket{ge}=\begin{pmatrix}
0\\
0\\
1\\
0
\end{pmatrix},
\ket{ee}=\begin{pmatrix}
0\\
0\\
0\\
1
\end{pmatrix},
\label{eq:Kroneckerbasis}
\end{equation}
where the first label in $\ket{ij}$ refers to Alice and the second label to Bob. 
It follows that
\begin{subequations}
\begin{align}
\mu_A(\tau) &= 
\left(
\begin{array}{c c c c}
0 & e^{-\ii \Omega_A \tau} & 0 & 0\\
e^{\ii \Omega_A \tau} & 0 & 0 & 0\\
0 & 0 & 0 & e^{-\ii \Omega_A \tau}\\
0 & 0 & e^{\ii \Omega_A \tau} & 0\\
\end{array}\right )
\ , 
\\
\mu_B(\tau) &= 
\left(
\begin{array}{c c c c}
0 & 0 & e^{-\ii \Omega_B \tau} & 0\\
0 & 0 & 0 & e^{-\ii \Omega_B \tau}\\
e^{\ii \Omega_B \tau} & 0  & 0 & 0\\
0 & e^{\ii \Omega_B \tau} & 0 & 0\\
\end{array}\right )
\ . 
\end{align}
\end{subequations}

We take the initial state of the Alice-Bob system to be the maximally entangled state  
$\ket{\psi_{\text{max}}}=\frac{1}{\sqrt2}\bigl(\ket{gg}+\ket{ee}\bigr)$, 
so that
\begin{align}
\label{eq:stateinitial}
\rho_{\text{d},0}=\proj{\psi_{\text{max}}}{\psi_{\text{max}}}=\frac12 \! \begin{pmatrix}
1 & 0 &0 &1\\
0 & 0 &0 &0\\
0 & 0 &0 &0\\
1 & 0 &0 &1\\
\end{pmatrix}
\, . 
\end{align}
In the final state $\rho_{\text{d},T}$ \eqref{eq:rho-d-T-expanded}, 
we separate the contributions to 
$\rho_{\text{d},T}^{(2)}$ as 
\begin{align}
\rho_{\text{d},T}^{(2)}=\lambda_A^2\rho_{AA}+\lambda_B^2\rho_{BB}+\lambda_A\lambda_B\rho_{AB}
\ , 
\label{eq:orderdec}
\end{align}
finding 
\begin{widetext}
\begin{align}\label{eq:matrices}
\rho_{AA}&=\frac{1}{2}\!\left(\!\!
\begin{array}{cccc}
   - 2 \Realpart (J^{AA}_{-+}) & 0 & 0 & - J^{AA}_{-+}- {J^{AA}_{+-}}^*  \\
 0 &  I^{AA}_{+-}&I^{AA}_{++}  & 0 \\
 0 & I^{AA}_{--} & I^{AA}_{-+}& 0 \\
 - J^{AA}_{+-} - {J^{AA}_{-+}}^* & 0 & 0 &    - 2 \Realpart (J^{AA}_{+-})\\
\end{array}\!\!
\right)\!,\; \rho_{BB}=\frac{1}{2} \!\left(\!\!
\begin{array}{cccc}
   - 2 \Realpart (  J^{BB}_{-+}) & 0 & 0 &- J^{BB}_{-+} - {J^{BB}_{+-}}^*   \\
 0 &  I^{BB}_{-+}& I^{BB}_{--}   & 0 \\
 0 & I^{BB}_{++}& I^{BB}_{+-} & 0 \\
  - J^{BB}_{+-} - {J^{BB}_{-+}}^*  & 0 & 0 &     - 2 \Realpart (J^{BB}_{+-}) \\
\end{array}\!
\!\right),
\notag
\\[1ex]
\rho_{AB}
&= \frac{1}{2} \! \left(
\begin{array}{cccc}
  - 2 \Realpart (J^{AB}_{--}+ J^{BA}_{--})& 0 \ \ & 0 & - J^{AB}_{--} -  J^{BA}_{--} - {J^{AB}_{++}}^* - {J^{BA}_{++}}^*  \\
 0 &   I^{AB}_{++}+I^{BA}_{--} \ \ & I^{AB}_{+-}+I^{BA}_{-+} & 0 \\
 0 & I^{AB}_{-+}+I^{BA}_{+-}\ \ &    I^{AB}_{--}+I^{BA}_{++} & 0 \\
- J^{AB}_{++}-  J^{BA}_{++} - {J^{AB}_{--}}^* - {J^{BA}_{--}}^*  & 0 \ \ & 0 &- 2 \Realpart (J^{AB}_{++}+ J^{BA}_{++})  \\
\end{array}
\right), 
\end{align}
\end{widetext}
where 
\begin{align}
I_{\epsilon, \delta}^{\nu,\eta}
& \!=\!\!
\int_{-\infty}^{\infty}\!\!\!\!\!\!\!\text{d}\tau \!\!\int_{-\infty}^{\infty}\!\!\!\!\!\!\!\text{d}\tau' 
\chi_\nu(\tau')\chi_\eta(\tau) \, 
e^{\ii (\epsilon\Omega_\nu \tau'\!+\delta\Omega_\eta\tau)} W[\sfx_\eta(\tau),\sfx_\nu(\tau')],
\notag
\\
J_{\epsilon, \delta}^{\nu,\eta}
& \!=\!\!
\int_{-\infty}^{\infty}\!\!\!\!\!\!\!\text{d}\tau \!\!\int_{-\infty}^{\tau}\!\!\!\!\!\!\!\text{d}\tau' 
\chi_\nu(\tau)\chi_\eta(\tau') \, 
e^{\ii (\epsilon\Omega_\nu \tau+\delta\Omega_\eta\tau')} W[\sfx_\nu(\tau),\sfx_\eta(\tau')]. 
\label{eq:IJ-int-def}
\end{align}

Finally, we characterise the entanglement in the Alice-Bob final 
state $\rho_{\text{d},T}$ by the negativity $\mathcal{N}$~\cite{Negat}. 
For a two-qubit system this monotone provides a strict criterion of 
entanglement in the sense that it vanishes 
if and only if a state is separable. 
Working perturbatively to order~$\lambda^2$, 
the negativity can be computed in a straightforward way from 
\eqref{orders} and 
\eqref{eq:Kroneckerbasis}--\eqref{eq:IJ-int-def}.

\begin{figure}
\includegraphics[width=0.48\textwidth]{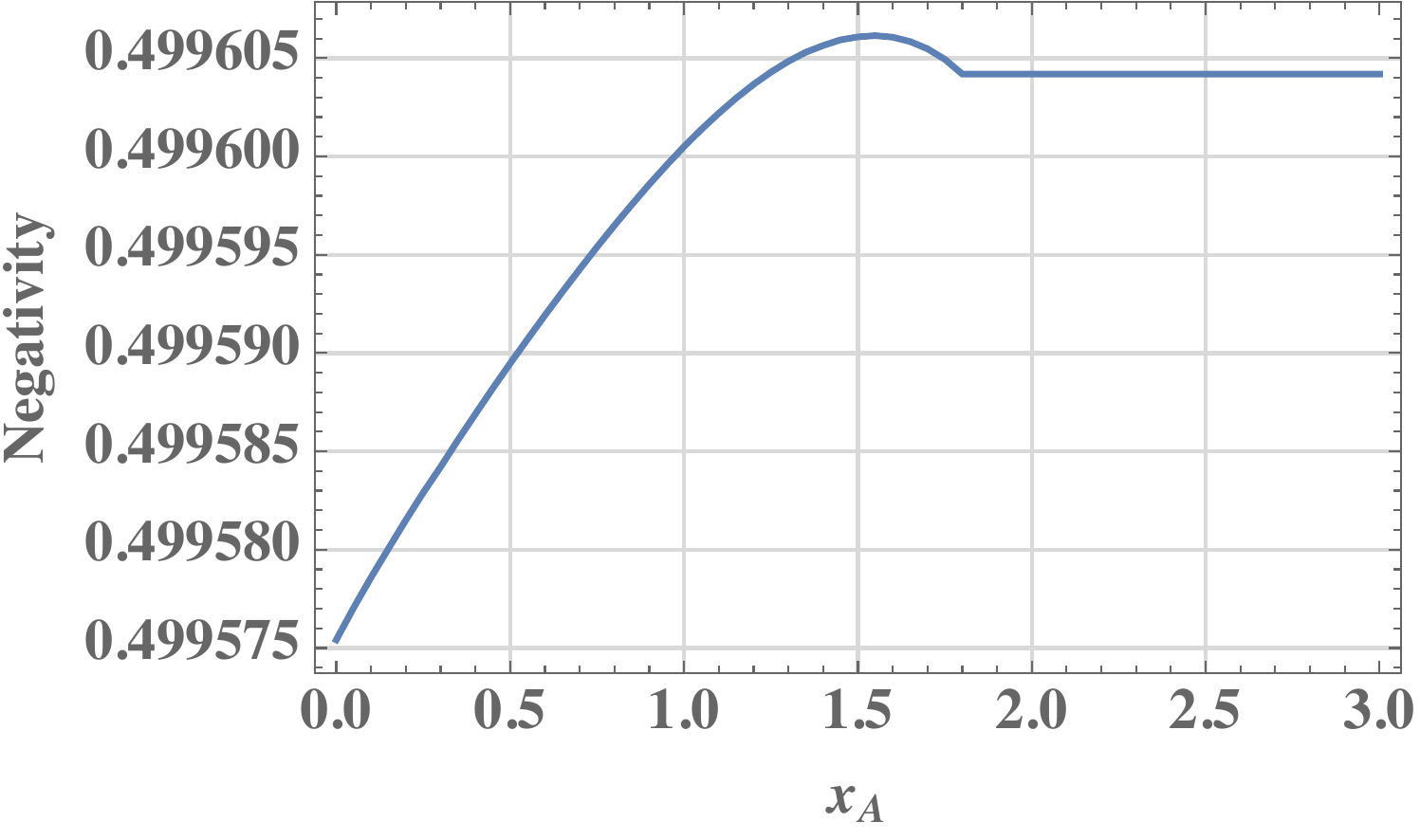}
\caption{Negativity $\mathcal{N}$ 
of the Alice-Bob final state as a function of $x_A$ with the detector trajectories 
shown in Figure \ref{fig:stdiagram} and with the sharp switching 
functions~\eqref{eq:chi-sharp}, 
for 
$R=4$, $T=1.8$, $\Omega_A = \Omega_B = 1$, 
and $\Lambda = 10^{-2}\ll\Omega_\nu$ and $\lambda_A =\lambda_B=0.01$. 
Before the interaction $\mathcal{N}=\frac12$, 
and the evolution causes the entanglement to degrade. 
When $x_A>T$, Alice does not cross the firewall during her detector's 
operation and the entanglement degradation is identical to that in Minkowski vacuum. 
When $x_A < T$, Alice does cross the firewall, 
and the entanglement degradation depends non-monotonically on~$x_A$. 
Note that the leading order contribution to 
$\mathcal{N}$ is a homogeneous quadratic in $(\lambda_A, \lambda_B)$:
an overall scaling of the two coupling constants 
(within the perturbative approximation)
leaves the plot invariant up to a rescaling of the vertical axis.}
\label{fig:sudden}
\end{figure}

\textit{\textbf{Results.---}}
With the detector trajectories shown in Figure~\ref{fig:stdiagram}, 
we first consider switching functions with a sharp switch-on and switch-off, 
\begin{align}
\chi_A(\tau) = \chi_B(\tau) = \Theta(\tau) \Theta\bigl(1 - (\tau/T)\bigr)
\ , 
\label{eq:chi-sharp}
\end{align}
where $\Theta$ is the Heaviside function. 
Figure \ref{fig:sudden} shows a representative plot of the 
negativity as a function of $x_A$ with the other parameters fixed. When $x_A >T$, 
Alice does not fall through the firewall during the operation of 
the detectors (see Fig.~\ref{fig:stdiagram}) 
and the entanglement degradation is just that in Minkowski vacuum~\cite{Reznik2005}, 
independent of~$x_A$. When $x_A <T$, Alice's falling through the firewall does affect the negativity. 
Two outcomes are apparent from the figure. 

First, the firewall effect on the negativity depends continuously on 
$x_A$ and remains small in magnitude: the firewall 
does \emph{not\/} wash up the Alice-Bob correlations 
as might have been expected from the gravitational firewall debate 
\cite{Mathur:2009hf,braunstein-et-al,Almheiri:2012rt,Susskind:2013tg,Almheiri:2013hfa,Page:2013mqa,marolf-polchinski-duality-bhint,Hotta:2013clt,Almheiri:2013wka,Harlow:2014yka,Hotta:2015yla}.
As a technical point, we note that the smallness of the effect gives confidence
in the reliability of our perturbative analysis. 

Second, over most of the parameter range the firewall enhances 
the degradation of Alice-Bob entanglement, 
compared with the degradation in Minkowski vacuum. 
This is what one might have 
expected from the gravitational firewall 
debate 
\cite{Mathur:2009hf,braunstein-et-al,Almheiri:2012rt,Susskind:2013tg,Almheiri:2013hfa,Page:2013mqa,marolf-polchinski-duality-bhint,Hotta:2013clt,Almheiri:2013wka,Harlow:2014yka,Hotta:2015yla}.
However, if Alice crosses the firewall shortly before turning 
her detector off, the effect is the opposite: in this case 
the firewall helps Alice and Bob maintain their entanglement. 
Developing a qualitative explanation for this phenomenon 
could be an interesting challenge. 

\begin{figure}
\includegraphics[width=0.48\textwidth]{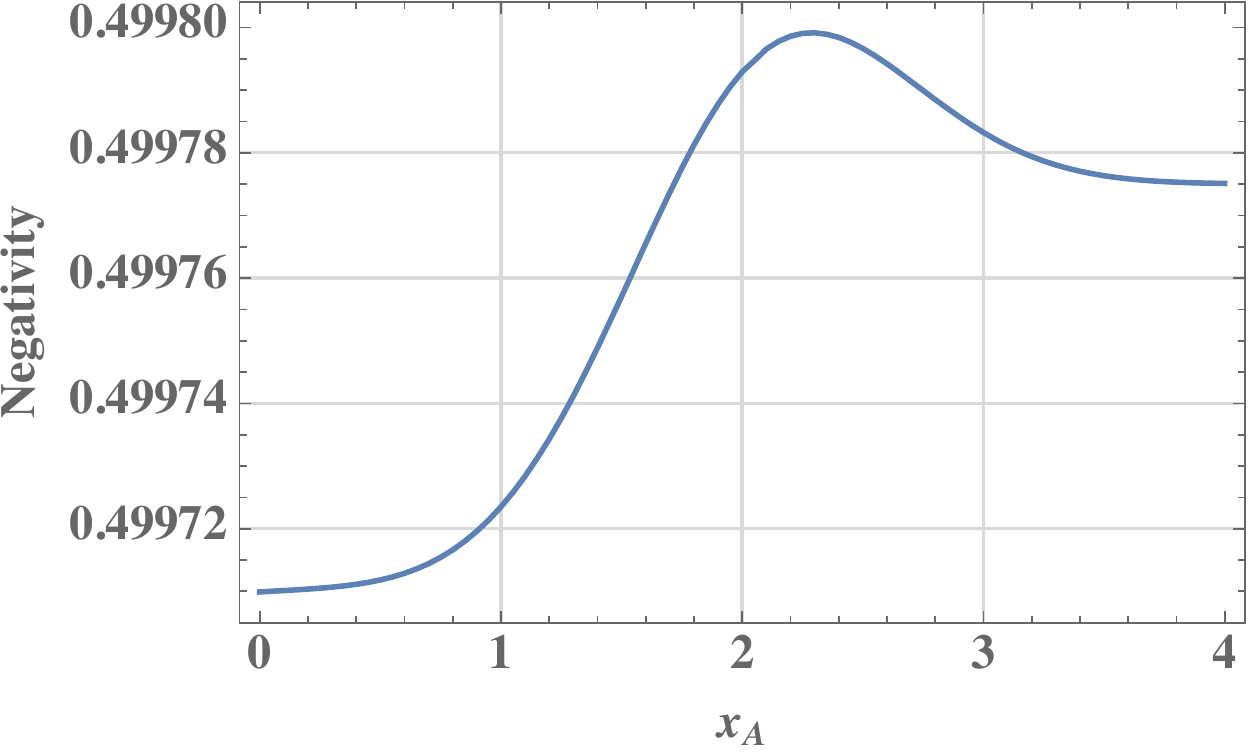}
\caption{As in Figure \ref{fig:sudden} 
but with the Gaussian switching functions~\eqref{eq:Gaussian}, 
with $\sigma=1$ and $t_0=2$, so that the Gaussian switching provides a 
smooth approximation to the sharp switching of Figure~\ref{fig:sudden}. 
The firewall effect on Alice's detector is now significant for $x_A \lesssim 3$. 
The qualitative properties are as in Figure~\ref{fig:sudden}, 
including the non-monotonicity in~$x_A$.}
\label{fig:gauss}
\end{figure}

One might suspect some of the properties of the graph in Figure 
\ref{fig:sudden} to be specific to, and perhaps artefacts of, the 
sharp switch-on and switch-off. To alleviate this suspicion, 
Figure \ref{fig:gauss} shows results from a similar analysis with
Gaussian switching functions, 
\begin{align}
\chi_A(\tau)=\chi_B(\tau)
= 
e^{-(\tau-\tau_0)^2/\sigma^2}
\ , 
\label{eq:Gaussian}
\end{align}
where the parameters $\tau_0$ and $\sigma$ are chosen as described in the 
figure caption to provide a smooth approximation to the sharp 
switching of Figure~\ref{fig:sudden}. 
The detectors now operate for $-\infty < \tau < \infty$, 
but the tails of the Gaussians are so small that this 
noncompact support of the Gaussian does not bring in new complications. 
The curve in Figure \ref{fig:gauss} is smoother but retains the qualitative features, 
including a regime where the firewall allows Alice and Bob maintain 
their entanglement better than in Minkowski vacuum. 
The conclusions drawn above from the sharp switching results 
hence apply also to the Gaussian switching. 

Generalising our analysis from $1+1$ to $3+1$ dimensions would require
new technical input at two steps. First, the firewall Wightman
function must be evolved from the initial data at $t=0$ using the
$(3+1)$-dimensional 
field commutator. 
Second, recall that while the
transition probability of a pointlike Unruh-DeWitt detector is well
defined in $3+1$ dimensions (an account that includes the switching
effects can be found in~\cite{Satz:2006kb,Louko:2007mu}), the
evolution of the full density operator is singular even in Minkowski
vacuum, as seen from \eqref{eq:IJ-int-def} and from the nonintegrable
coincidence limit singularity of the $(3+1)$-dimensional Wightman
function \cite{Kay:1988mu,Decanini:2005gt}.
A~decoherence analysis would hence require an additional 
regularisation of the detector model, for example by a spatial smearing, 
 a standard and 
well documented procedure \cite{Unruh,Schlicht:2003iy,Langlois:2005nf,Louko:2006zv}.
However, crucially, the structure of \eqref{eq:matrices} remains exactly the
same in all dimensions. It is hence difficult to see how any
reasonable input at either of these technical steps could lead to
drastically enhanced singularities in the evolution across the
firewall. In particular, correlations between a firewall-crossing
detector and the outside world should still undergo only a modest
change.

\textit{\textbf{Conclusions.---}} Our main conclusion runs contrary to
the vision of a firewall as a violently singular
surface~\cite{Almheiri:2012rt}: the Rindler
firewall has only a modest effect on the entanglement between two
inertial Unruh-DeWitt detectors when one of the detectors crosses the
firewall. There is even a parameter range in which the firewall
slows down the entanglement degradation, compared
with the degradation that takes place in Minkowski
vacuum.

Given that the Rindler firewall models the quantum field theory correlations
in a black hole firewall~\cite{Louko2014}, 
our results suggest that a similar
conclusion should hold for black hole 
firewalls at the early stages of the Hawking evaporation where the
gravitational backreaction on the metric is not yet significant. As
the Unruh-DeWitt detector captures the essential features of the
interaction between atoms and the electromagnetic
field~\cite{Wavepackets,Alvaro}, the conclusion should further extend
to systems of matter of which we and our experimental apparatus are
built.

In summary, the key message of this letter is that 
we cannot think of a young firewall as a surface of cataclysmic
events that erases all information about matter that crosses the firewall. 
If the matter is correlated with the
outside world, these correlations will not be significantly altered by
the crossing. We may not know why the chicken crossed the young firewall,
but it did get to the other side, with most of its memories intact.

More broadly, our results push the burden of proof of the 
firewall's ability to resolve the black hole information paradox into 
the regime in which the detailed late time gravitational structure of 
the firewall is crucial. This regime is at present 
conspicuously poorly understood, and not exempt of problems \cite{Abramowicz}
(a rare exception is a dilaton gravity model 
\cite{Almheiri:2013wka} in which 
the paradox turns out to be resolved by a remnant 
rather than by a firewall). 
While our results do not settle the viability of the firewall argument, 
they do identify the 
arena in which the viability will be settled.

\textit{\textbf{Acknowledgments.---}} The authors thank Don Marolf for asking how a detector
responds in the Rindler firewall state considered in this Letter and
in Ref.~\cite{Louko2014}. 
JL thanks EMM and Achim Kempf for hospitality at the 
University of Waterloo. 
EMM was supported by the NSERC Discovery programme. JL was supported in part by STFC 
(Theory Consolidated Grant ST/J000388/1).

\bibliography{biblio}

\end{document}